\documentclass{aa}  

\usepackage[breaklinks=true]{hyperref} %% to avoid \citeads line fills
\bibpunct{(}{)}{;}{a}{}{,}             %% natbib format for A&A and ApJ
 
\usepackage{graphicx}
\usepackage{amsmath}		% Advanced maths commands
\usepackage{amssymb}	        % Extra maths symbols
\usepackage{subfigure}
\usepackage{longtable}
\usepackage{tabularx, blindtext}
\usepackage[varg]{txfonts}
\usepackage{lmodern}

\begin{document} 

\title{The \textit{Hubble Space Telescope} UV Legacy Survey \\ of Galactic Globular Clusters}
\subtitle{XXII. Relative ages of multiple populations in five Globular Clusters.}
\titlerunning{Relative ages of multiple populations in five Globular Clusters.}

   \author{F. Lucertini\inst{1}, D. Nardiello\inst{2,3}, G. Piotto\inst{4,3}}

   \institute{Departamento de F\'isica, Faculdad de Ciencia Exactas, Universidad Andr\'es Bello, Fern\'andez Concha 700, Santiago, Chile
	\and   	
   	Aix Marseille Univ, CNRS, CNES, LAM, Marseille, France
   	\and
   	Istituto Nazionale di Astrofisica - Osservatorio Astronomico di Padova, Vicolo dell’Osservatorio 5, Padova, IT-35122
   	\and 
   	Dipartimento di Fisica e Astronomia "Galileo Galilei", Universit\'a di Padova, Vicolo dell’Osservatorio 3, Padova IT-35122
   	}

   \date{Received XXX ; accepted YYY}

% \abstract{}{}{}{}{} 
% 5 {} token are mandatory
 
  \abstract
   {}
   {We present a new technique to estimate the relative ages of multiple stellar populations hosted by five globular clusters: NGC\,104 (47\,Tuc), NGC\,6121 (M4), NGC\,6352, NGC\,6362 and NGC\,6723.}
   {We used the catalogs of the database ``HST UV Globular cluster Survey (HUGS)'' to create color-magnitude and two-color diagrams of the Globular Clusters. We identified the multiple populations within each globular cluster, and we divided them into two main stellar populations: POPa or first generation (1G) and POPb, composed of all the successive generations of stars. The new technique allows us to obtain an accurate estimate of the relative ages between POPa and POPb.}
   {The multiple populations of NGC\,104 and NGC\,6121 are coeval within 220 Myr and 214 Myr, while those of NGC\,6352, NGC\,6362 and NGC\,6723 are coeval within 336 Myr, 474 Myr and 634 Myr, respectively. These results were obtained combining all the sources of uncertainties.}
   {}

   
   \maketitle

%
%________________________________________________________________

\section{INTRODUCTION}
The concept that Galactic Globular Clusters (GCs) host multiple stellar populations 
is supported by an overwhelming amount of observational facts, and accepted by the 
astronomical community. However, the origin and the time-scales for the formation and 
the evolution of multiple populations (MPs) are still under debate and study. 
The most accredited scenarios support that MPs phenomenon in GCs is due to
multiple events of star formation.  These formation scenarios consider
the existence of a first generation (1G) characterized by stars having
chemical properties similar to that of the interstellar medium out of
which they formed, and a second generation (2G),
formed from the material processed by 1G stars.  
Among the proposed alternatives to explain 2G stars,
the intermediate mass Asymptotic giant branch
(AGB) scenario \citep{2002A&A...395...69D} predicts that 2G stars were born
from the AGB ejecta within 100\,Myr \citep{2012MNRAS.423.1521D}.
On the other hand, \cite{2009A&A...507L...1D} proposed that 2G population was born 
from processed low-velocity material ejected by massive binaries in a time-scale of $\sim$1\,Myr. 
Suggesting supermassive stars with $\mathcal{M} \sim 10^4\, M_{\sun}$ as 1G stars, 
\cite{2014MNRAS.437L..21D} demonstrated that 2G stars from in $10^5$ yr.
Therefore, the relative age of MPs can provide clues about formation and evolution of 
these populations and can also discriminate among the different scenarios proposed. \\
In the last years, the investigation of MPs was expanded to GCs outside the Milky Way (MW). 
The fact that not only our Galaxy hosts MPs allows us to compare GCs formed in systems with different star formation histories, providing clues on MPs phenomenon. Moreover, it turns out that the environment is not one primary requirement of MPs formation.
\cite{2019MNRAS.485.3076N} analyzed the stellar populations within the GC Mayall II (G1), located in the halo of the nearby Andromeda galaxy.
Several works focused on the discovery that intermediate age (1-2 Gyr) clusters in the Magellanic Clouds show extended main sequence turn off and splitted main sequence (MS) (\citealt{2019MNRAS.489L..97S}, \citealt{2019MNRAS.486.5581G}, \citealt{2019MNRAS.487.5324M}).
\cite{2009MNRAS.398L..11B} proposed the rotational velocity of stars with masses between 1.2-1.7 $M_{\sun}$ to explain the presence of double MS.
The age is another fundamental factor, since the lower age limit of a star cluster to show MPs is 2 Gyr (\citealt{2017MNRAS.468.3150M}).
The MPs in the young star cluster NGC\,1978 in the Magellanic Clouds are coeval within 1$ \pm$20 Myr (\citealt{2018MNRAS.477.4696M}). Similarly, \cite{2020MNRAS.493.6060S} found that the star cluster NGC\,2121 hosts coeval MPs within 6$\pm$12 Myr.
These results reveal that young GCs put tighter constraints than older ones on the MPs formation timescale.\\
The Treasury program ``The Hubble Space Telescope (HST) UV Legacy Survey of 
Galactic Globular Clusters'' (GO-13297, PI: Piotto, \citealt{2015AJ....149...91P}) provided us 
with an unprecedented UV dataset of more than 50 GCs. 
The aim of the project was the photometric characterization of the MPs in GCs by combining 
the new UV data with optical HST data from the program ``ACS Survey of GCs" 
(GO-10775, PI: Sarajedini, \citealt{2006AAS...20910009S}).
Based on this dataset, \cite{2015MNRAS.451..312N} developed a new procedure 
for evaluating the relative age of MPs within NGC\,6352, assuming the different 
populations in the cluster have the same metallicity.
Recently, \cite{2020ApJ...891...37O} employed a new code to provide a statistical 
fitting of isochrones to observed CMDs and to derive age differences 
between 1G and 2G in Bulge GCs.\\
In this work, we present a new technique developed to estimate
the relative age of MPs, using their main sequence turn off (MSTO)
as indicator of age. The GCs we considered in this work are five: 
NGC\,104 (47\,Tuc), NGC\,6121 (M4), NGC\,6352, NGC\,6362 and NGC\,6723. 
A proper identification of 1G stars and those of subsequent generations is a fundamental  
point for the application of our method. For this reason, we selected these objects because 
they show well-separated multiple sequences in their UV CMDs.
The last three GCs  of our sample were previously analyzed by \cite{2015MNRAS.451..312N} 
and \cite{2020ApJ...891...37O}, providing a direct evaluation of the method reliability.\\
The paper is organized as follows. Data reduction and analysis are briefly described in Section 2.
In section 3, we show the procedure adopted to characterize MPs within the GCs.
In section 4, the new method is explained. The results and comparison with literature are 
discussed in section 5. Summary and conclusions follow in Section 6.

%______________________________________________________________________

\section{OBSERVATION AND DATA REDUCTION}
In this work we used the catalogs obtained in the project ``HST UV Globular cluster Survey" 
(HUGS\footnote{https://archive.stsci.edu/prepds/hugs/\\ DOI: 10.17909/T9810F}, \citealt{2018MNRAS.481.3382N}).  
For a detailed description of the data-reduction pipeline used to obtain these catalogs, 
we refer the reader to \cite{2017ApJ...842....6B} and \cite{2018MNRAS.477.2004N}.  
The catalogs contain the positions of the stars, the magnitude in five filters 
(F275W, F336W, F438W, F606W, and F814W), and quality parameters such as 
the photometric errors in a filter X ($\sigma_{\rm X}$), the
quality-of-fit (QFIT), and the shape of the source (SHARP).\\
To analyze the MPs within GCs, we selected well measured stars
on the basis of these parameters, as done by \cite{2018MNRAS.477.2004N}. 
Briefly, we divided the sample of stars in a given filter X into bins of
0.5 magnitudes, and we calculated the 3$\sigma$-clipping median values
in each bin. We interpolated these median values using a spline,
and we rejected all the stars that are 3$\sigma$ above (in the case of $\sigma_{\rm X}$ )
or below (in the case of QFIT) the median parameters. Stars with
-0.2$<$SHARP$<$0.2 were considered as well measured. All
stars that satisfy the three conditions were selected and used
during the analysis data.\\
We used the procedure described in \cite{2012A&A...540A..16M} to correct 
the magnitudes for differential reddening.  
In the case of NGC\,104, NGC\,6362 and NGC\,6723, the results obtained 
from this procedure do not lead either to a significant correction or  to a 
considerable improvement of the CMDs.  
For this reason, we decided to continue the analysis without taking into 
consideration the differential reddening correction for these objects.

%_________________________________________________________________________________________
\begin{figure}
  \centering
	\includegraphics[trim= 0.8cm 5.5cm 1cm 2.8cm, clip,width=0.4\textwidth]{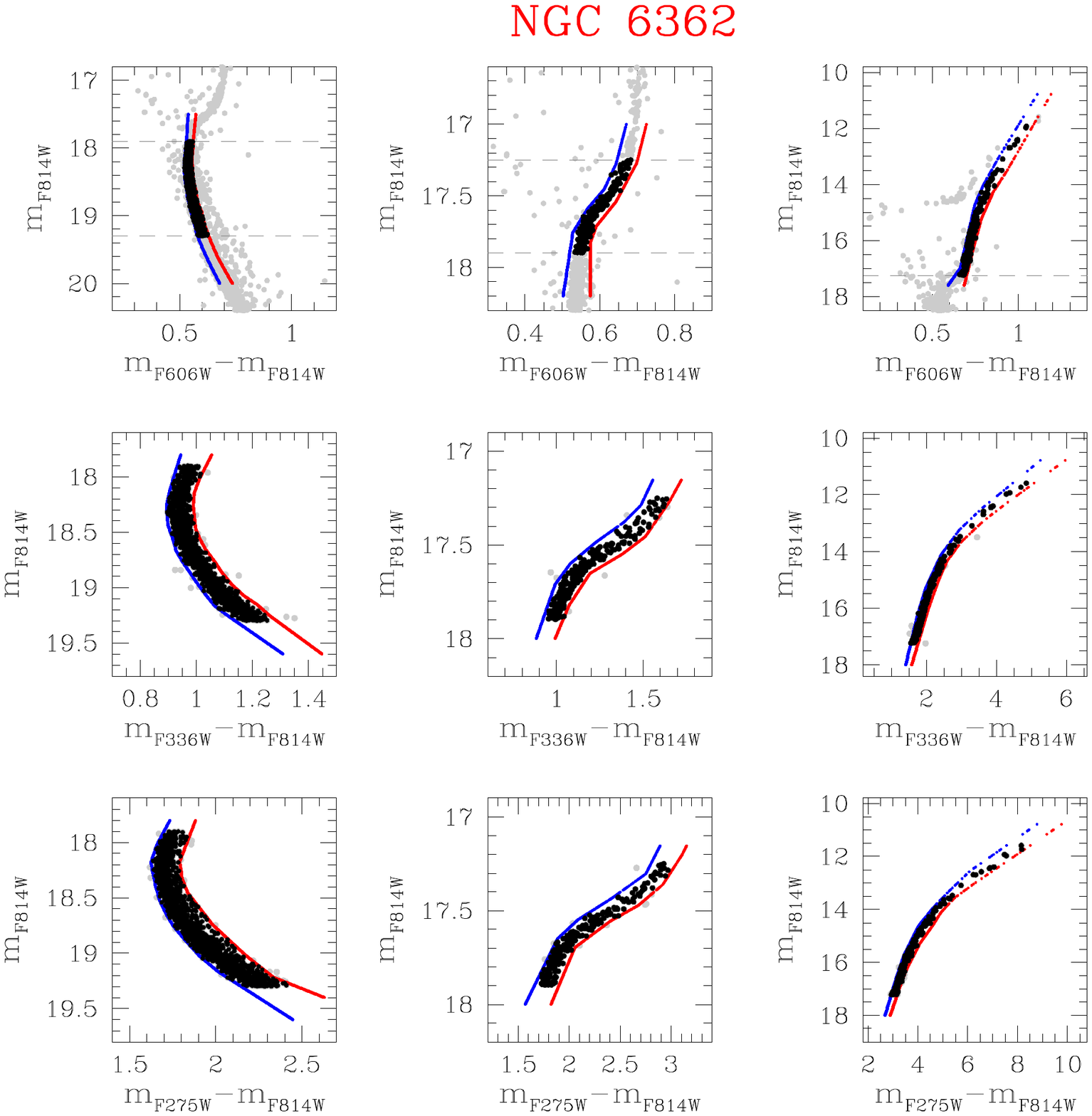}
	\caption{Procedure adopted to select MS (left-hand panels), SGB (middle panels) and RGB stars (right-hand panels) in the $m_{\rm F814W}$ versus $m_{\rm F606W}-m_{\rm F814W}$ (top panels), $m_{\rm F814W}$ versus $m_{\rm F336W}-m_{\rm F814W}$ (middle panels) and $m_{\rm F814W}$ versus $m_{\rm F275W}-m_{\rm F814W}$ (bottom panels) CMDs of NGC\,6362. The hand drawn fiducial lines of each evolutionary phase are reported in blue and red. Grey and black points represent rejected and selected stars, respectively. The $m_{\rm F814W}$ magnitude ranges where MS, SGB and RGB stars were selected are shown by horizontal dashed lines in the top panels.}
	\label{sel}	
\end{figure}

\begin{figure}
  \centering
	\includegraphics[trim = 0.8cm 5.5cm 1cm 2.8cm, clip, width=0.45\textwidth]{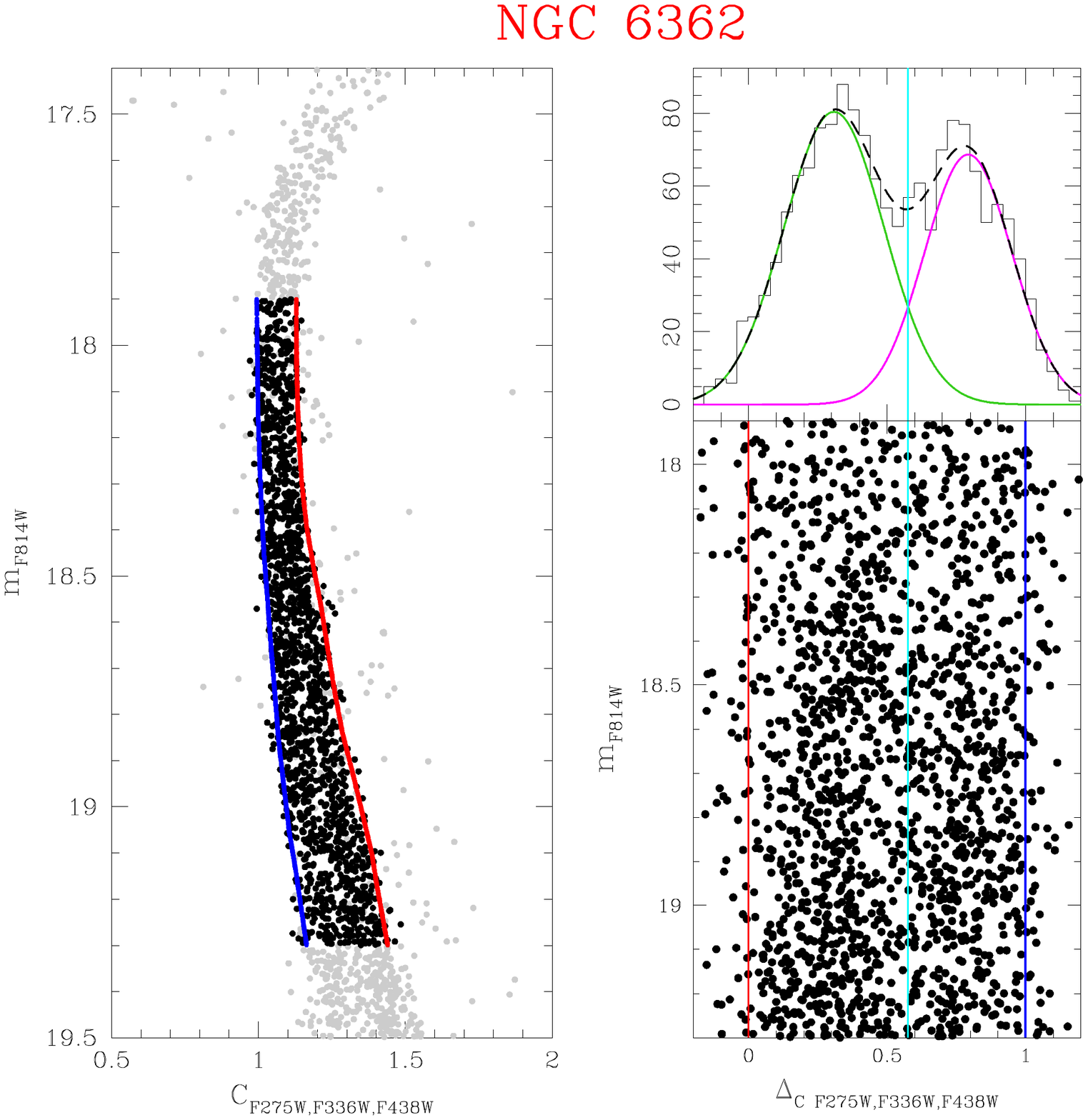}
	\caption{Statistic separation of MS stars of the GC NGC\,6362. In the $m_{\rm F814W}$ vs $C_{\rm F275W,F336W,F438W}$ pseudo-CMD are reported well measured (grey) and MS selected (black) stars. The blue and red fiducial line represent the 4$^{th}$ and 96$^{th}$ percentiles of the distribution in color. The left bottom panel shows the verticalized MS region. The histogram represents the stars distribution of the verticalized pseudo-CMD. The vertical cyan line is used to discriminate between MS of POPa (left hand-side) and Ms of POPb (right hand-side).}
	\label{stat}	
\end{figure}

\begin{figure*}
  \centering
	\includegraphics[trim= 0.8cm 5.5cm 1cm 2.8cm, clip, width=0.7\textwidth]{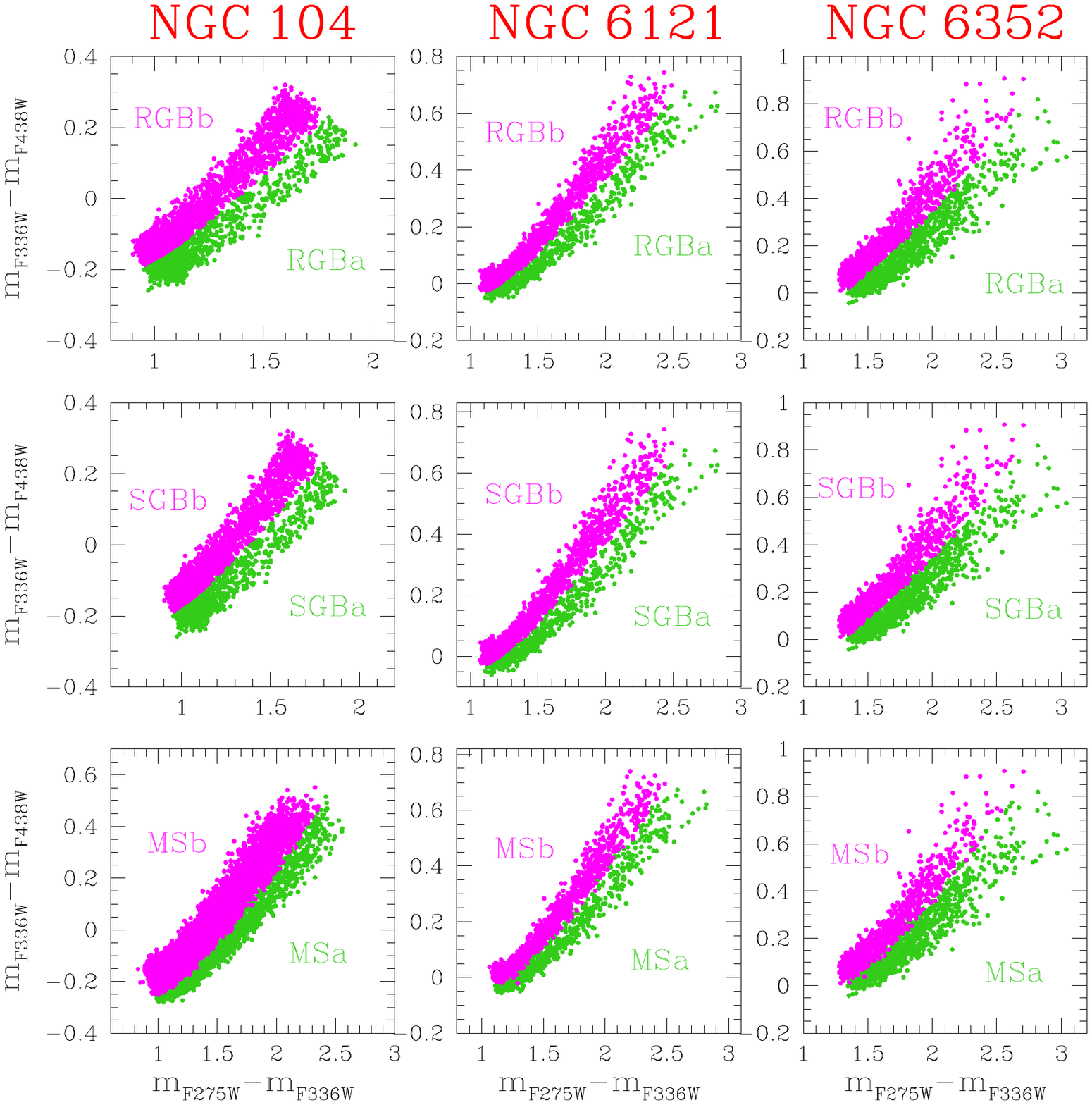}
	\caption{The $m_{\rm F336W}-m_{\rm F438W}$ versus $m_{\rm F275W}-m_{\rm F336W}$ two-color diagram of MS (bottom panels), SGB (middle panels) and RGB (top panels) stars of POPa (green) and POPb (magenta) belonging to NGC\,104 (left-hand panels), NGC\,6121 (center panels), NGC\,6352 (right-hand panels).}
	\label{tcd1}	
\end{figure*}

\begin{figure*}
  \centering
	\includegraphics[trim= 0.7cm 5.5cm 6cm 2.8cm, clip, width=0.5\textwidth]{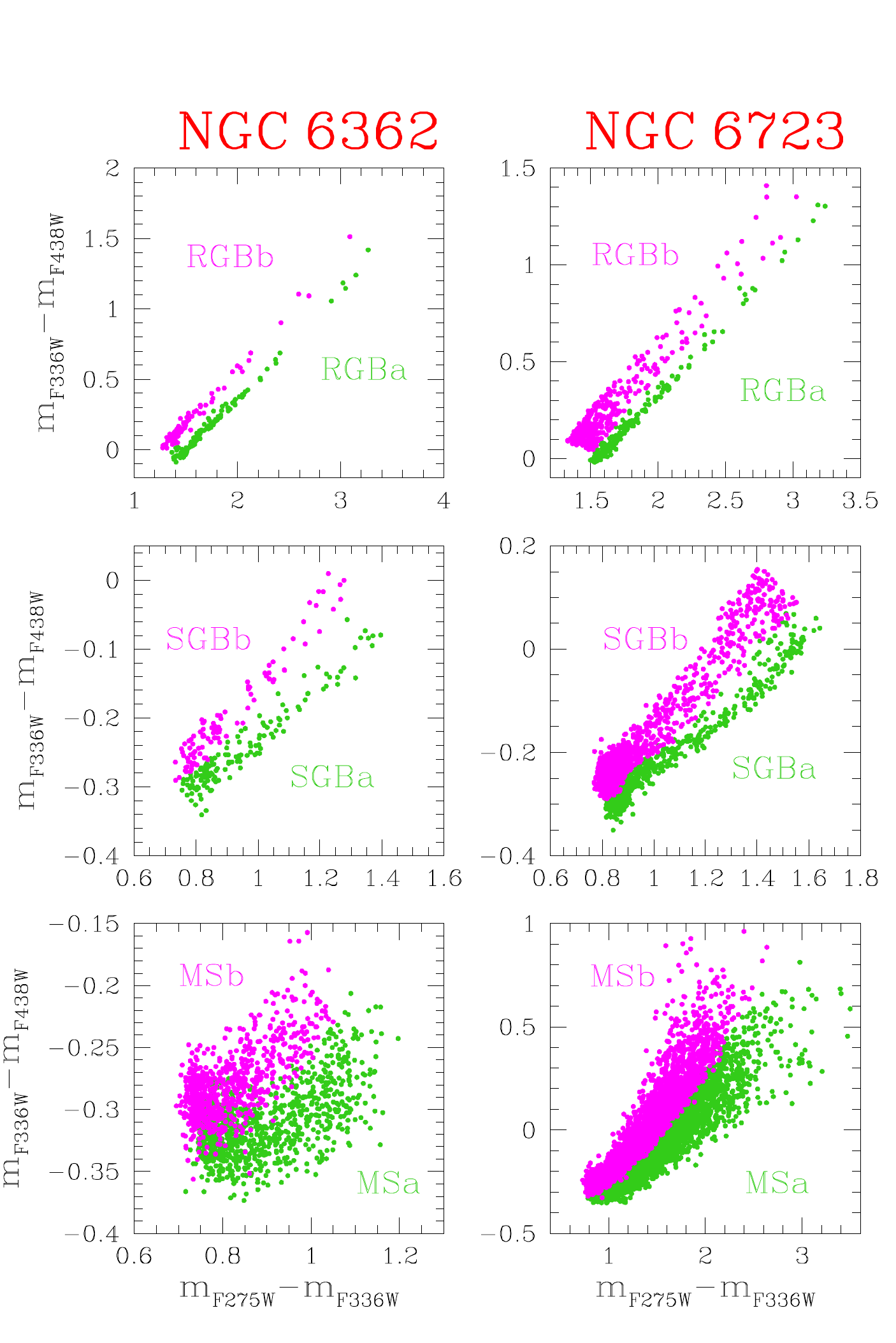}
	\caption{As in Fig.\ref{tcd1} for the GCs NGC\,6362 (lef-hand panels) and NGC\,6723 (right-hand panels).}
	\label{tcd2}	
\end{figure*}

\section{MULTIPLE STELLAR POPULATIONS WITHIN GLOBULAR CLUSTERS}
Since optical filters are less sensitive to light elements variations, such as C, N and O, they allow us to identify MPs in CMDs only when metallicity and/or C+N+O content changes among the cluster stars (NGC\,1851, 
\citealt{2012MSAIS..19..173M}; M22 \citealt{2012MSAIS..19..173M} and M2, 
\citealt{2015MNRAS.447..927M}). 
Each cluster analyzed in this work hosts stars having all the same metallicities and C+N+O content, and in this manuscript we refer to multiple populations as the phenomenon due to the variation of light elements (C,N,O) among the cluster stars. In this context, the UV HST filters F275W, F336W and F438W are efficient for
the identification of MPs in GCs,  because they are sensitive to the variations of the molecular bands of  OH, NH, CH and CN.\\
Taking advantage of this possibility, we selected stars on the MS, sub-giant branch (SGB) and red-giant branch (RGB) and rejected those that are not on these three evolutionary phases, in three steps.
The GC NGC\,6362 is taken as an example in Fig. \ref{sel}.
First, we selected the stars in the $m_{\rm F814W}$ versus $m_{\rm F606W} - m_{\rm F814W}$ CMD (top panels).
We draw by hand two fiducial lines: one on the blue and one on the red side of each evolutionary phase.
We rejected the stars lying on the left-hand side and on the right-hand side of the blue and red 
fiducial lines, respectively (grey points). The stars lying between the two fiducial lines and in a specific $m_{\rm F814W}$ range were selected (black points).
Then, we followed the same procedure in the $m_{\rm F814W}$ versus 
$m_{\rm F336W} - m_{\rm F814W}$ CMD, considering the previously selected stars (middle panels).
In the final step, we selected in the same way the stars in the 
$m_{\rm F814W}$ versus $m_{\rm F275W} - m_{\rm F814W}$ CMD (bottom panels).\\
For the final goal of this work, it is not necessary to perfectly separate all the MPs hosted by the GCs.
Indeed, we are interested to identify a statistic amount of each population stars. 
For this reason, we will divide the MPs of each GC into two main stellar populations.
In particular, from now on, we will refer to population-a (POPa) as the first generation stars (Na- and N- poor, and O- and C-rich) and to population-b (POPb) as successive generations.\\
Since our technique is based on the use of the MSTO color, we need to identify the bulk of POPa and POPb stars on the MS of each cluster. We applied a statistical identification based on the use of the pseudo-color $C_{\rm F275W, F336W, F438W}=(m_{\rm F275W}-m_{ \rm F336W})-(m_{\rm F336W}-m_{\rm F438W})$, that allows us to maximize the separation between the different populations (see \citealt{2013ApJ...767..120M}). Figure \ref{stat} shows the procedure adopted.
Well measured stars are reported in gray, while black points are the selected MS stars.
The blue and red fiducial lines are the 4$^{th}$ and 96$^{th}$ percentiles of the color distribution of MS stars.
In order to obtain the fiducial lines, we divided the MS into a set of F814W magnitude bins of size $\delta m$, subdivided
into sub-bins of $\delta m/3$. Moving iteratively of sub-bin, we calculated the 4$^{th}$, 96$^{th}$ percentiles of the color distribution and the mean F814W magnitude within each bin. The points were smoothed with a boxcar averaging filter and interpolated with splines.
In order to verticalize the pseudo-CMD, we transformed the $C_{\rm F275W, F336W, F438W}$ color using the following equation:
\begin{gather}
\mathrm{\Delta_{C F275W, F336W, F438W}} = \frac{\mathrm{Y_{fiducialR}} - \mathrm{Y}}{\mathrm{Y_{fiducialR}}-\mathrm{Y_{fiducialB}}}
\end{gather}
\noindent
where $Y=C_{\rm F275W,F336W,F438W}$, "fiducialR" and "fiducialB" are the red and blue fiducial lines. 
The distribution of stars in the $m_{\rm F814W}$ vs $\Delta_{\rm C F275W,F336W,F438W}$ diagram is represented by the histogram. Fitting a bimodal gaussian profile, we identified the $\Delta_{\rm C F275W,F336W,F438W}$ value useful to separate the stars into the MPs which they belong to. The stars on the left and right hand-side of the cyan vertical line were classified as MSa and MSb, respectively.
The two-color diagram $m_{\rm F336W}-m_{\rm F438W}$ versus 
$m_{\rm F275W}-m_{\rm F336W}$ of MS stars of each GC are shown in Figures \ref{tcd1} and\ref{tcd2}. 
We built the same diagrams for SGB and RGB stars of each GC. Since in the two-color diagrams of SGB and RGB stars two sequences are always clearly visible, in this case we divided the MPs drawing by hand a continuous line. 
We identified stars below and above the line as stars belonging to POPa (green) and stars belonging to POPb (magenta), respectively.
The samples of stars that belong to POPa and POPb of each GC analyzed in this work are formed 
by the combination of MS, SGB and RGB selected stars.

%_____________________________________________________________________________________________

\begin{figure*}
\centering
	\includegraphics[trim= 1.5cm 5.5cm 1cm 2.8cm, clip,width=0.7\textwidth]{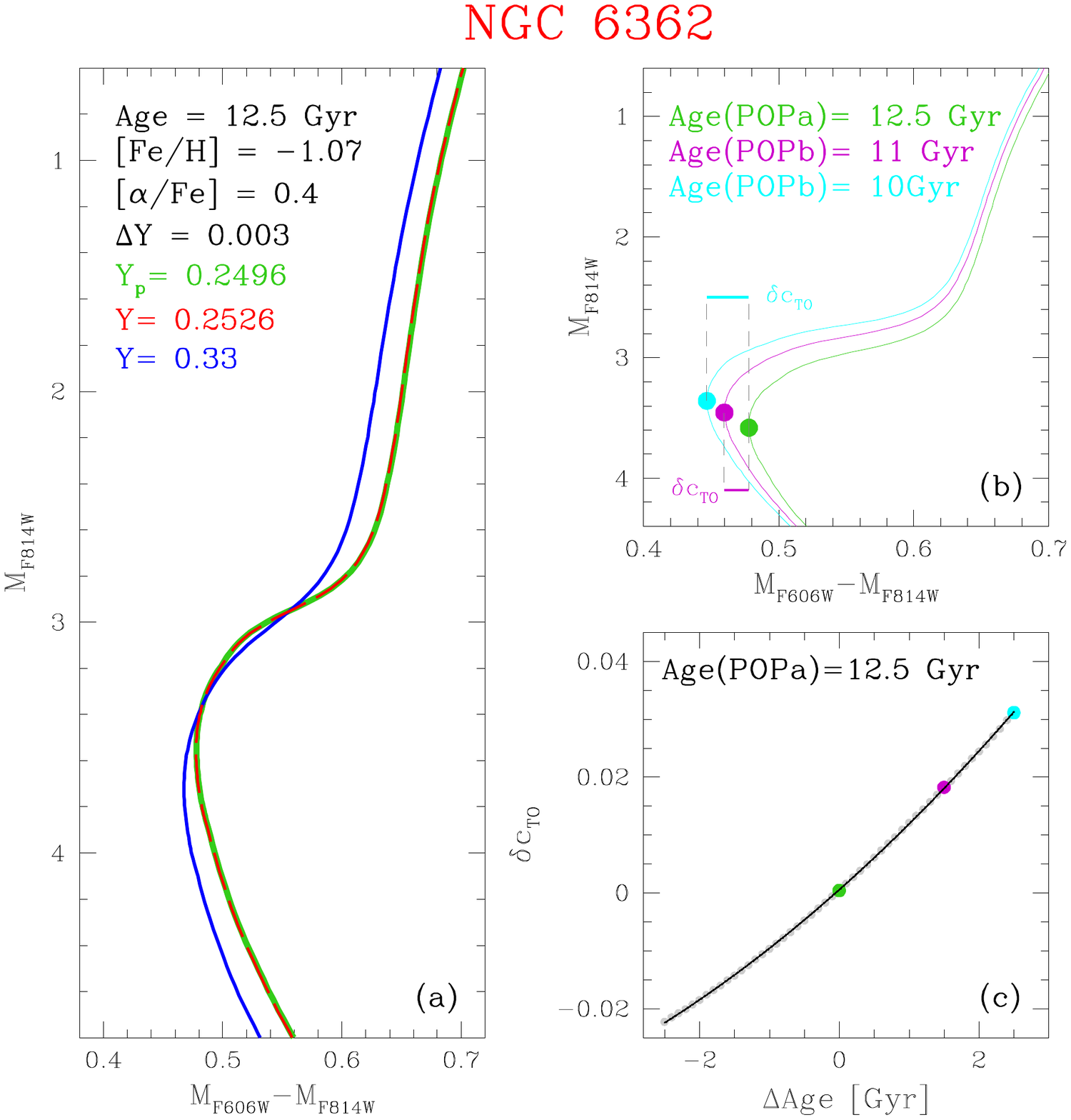}
	\caption{Procedure adopted to obtain the reference theoretical models for NGC\,6362. Panel (a) shows three isochrones of 12.5 Gyr, for [Fe/H]=-1.07, [$\alpha$/Fe]=0.4 and different helium content. The red isochrone with Y=0.2526 (helium-enriched by $\Delta$Y=0.003) is obtained interpolating the isochrone with primordial helium, Y$_p$=0.2496 (green), with the Y=0.33 helium-enriched one (blue). Panel (b) shows how the difference in color among the MSTOs, from which the $\delta c_{TO}$ is calculated. The green isochrone is the same as in panel (a) and represents POPa. The purple and cyan isochrones have the same chemical content as the red isochrone of panel (a), but with age of 11 Gyr and 10 Gyr, respectively. The corresponding $\delta c_{TO}$ of each case is a horizontal line. The theoretical model $\delta$c$_{TO}$ vs $\Delta$Age for NGC\,6362 is shown in panel (c). The purple and cyan points are the theoretical model values as obtained from the cases of panel (b).}
	\label{teo}
\end{figure*}

\section{RELATIVE AGE OF MPs WITHIN THE SELECTED GCs.}
The analysis of relative ages of MPs allows us to improve our
knowledge about the formation and evolution of the different
populations hosted by GCs.\\
In this work we propose a new technique to estimate the relative age
of the two main populations, POPa and POPb, hosted by five GCs: 
NGC\,104 (47\,Tuc), NGC\,6121 (M4), NGC\,6352, NGC\,6362, NGC\,6723. \\
In the previous section, we have taken advantage of the UV filters to
identify the two main populations hosted by our sample of GCs. 
In order to obtain the relative age between POPa and POPb we considered
them as simple stellar populations, and we used the optical 
$m_{\rm F814W}$ versus $m_{\rm F606W}-m_{\rm F814W}$ CMD because 
(in first approximation) these filters are not affected by light element variations.\\
We estimated the relative age of the two populations comparing the observed MSTO color with theoretical models.
Our technique is inspired by the horizontal method introduced by \cite{1999AJ....118.2306R}.
This method considers a point on the RGB and, the shape of this evolutionary phase mainly depends on the metallicity of the population.
In ``normal'' GCs the different MPs have the same metal content within the errors,
consequently, the point defined on the RGB will be almost the same.
In order to avoid the introduction of more photometric errors associated with the RGB color in the computation of relative ages, we used a method that does not take into account this point.
The innovation of our method resides on the MSTO colors difference, 
imposing strong constraints on metallicity and Helium content.
In this way, we achieve a differential measure of age that is independent from distance and reddening.

\begin{table*}
	\centering
	\caption{[Fe/H] (\citealt{2009A&A...505..117C} for NGC\,104, NGC\,2808, NGC\,6121, NGC\,6723; \citealt{2015MNRAS.451..312N} for NGC\,6352; \citealt{2017MNRAS.468.1249M} for NGC\,6362), [$\alpha$/Fe],  mean difference in helium $\Delta Y$ between POPa and POPb  \citep{2018MNRAS.481.5098M} and absolute age of POPa, Age(POPa), \citep{2010ApJ...708..698D}, considered to obtain the theoretical model of each GC.}
	\label{teomod}
	\small
	\centering
	\begin{tabular}{l|c|c|c|c}	
		\hline
		\textbf{Cluster} & \textbf{[Fe/H]} & \textbf{[$\alpha$/Fe]} & \textbf{$\Delta$Y} & \textbf{Age(POPa)} \\
		&&&&[Gyr] \\
		\hline
		NGC\,104& -0.76 $\pm$ 0.02 & 0.4 & 0.011 $\pm$ 0.005 &12.8 $\pm$0.5\\
		NGC\,6121& -1.18 $\pm$ 0.02 & 0.4 & 0.009 $\pm$ 0.006 &12.5 $\pm$0.5\\
		NGC\,6352& -0.67 $\pm$ 0.02 & 0.4 & 0.019 $\pm$ 0.014 & 13.0 $\pm$0.5\\
		NGC\,6362& -1.07 $\pm$ 0.05 & 0.4 & 0.003 $\pm$ 0.011  & 12.5 $\pm$0.5\\
		NGC\,6723& -1.10 $\pm$ 0.07  &0.4  & 0.005 $\pm$ 0.006 & 12.8 $\pm$0.5\\
		\hline		
	\end{tabular}
\end{table*}

\begin{table*}
	\centering
	\caption{Value of color and magnitude of MSTO and observed $\delta$c$_{TO}$ obtained for POPa and POPb within GCs}
	\label{MSTOGCs}
	\begin{tabular}{l|c|c|c|c|c}
		\hline
		\textbf{Cluster}& \multicolumn{2}{|c|}{\textbf{MSTO(POPa)}} &\multicolumn{2}{|c|}{\textbf{MSTO(POPb)}} & \textbf{$\delta$c$_{TO}$}\\
		& $m_{\rm F606W}-m_{\rm F814W}$ & $m_{\rm F814W}$ & $m_{\rm F606W}-m_{\rm F814W}$ & $m_{\rm F814W}$ & \\
		\hline
		NGC\,104 & 0.5368$\pm$0.0004 & 17.024$\pm$0.004& 0.5359$\pm$0.0002 & 16.995$\pm$0.002 & 0.0009$\pm$0.0005 \\
		NGC\,6121  &  0.904$\pm$0.001 & 15.67$\pm$0.01 & 0.9039$\pm$0.0008 & 15.904$\pm$0.008 & 0.0005$\pm$0.0015 \\
		NGC\,6352 &  0.768$\pm$0.001 & 17.765$\pm$0.009 & 0.763$\pm$0.001 & 17.925$\pm$0.009 & 0.005$\pm$0.001 \\
		NGC\,6362  & 0.540$\pm$0.001 & 18.200$\pm$0.008 & 0.539$\pm$0.001 & 18.175$\pm$0.009 & 0.001$\pm$0.001 \\
		NGC\,6723 &  0.5564$\pm$0.0005 & 18.273$\pm$0.005 & 0.5555$\pm$0.0004 & 18.309$\pm$0.004 & 0.0010$\pm$0.0007 \\
		\hline			
	\end{tabular}
\end{table*}

\begin{figure*}
	\centering
	\includegraphics[trim= 0.8cm 5.5cm 6cm 2.8cm, clip,width=0.6\textwidth]{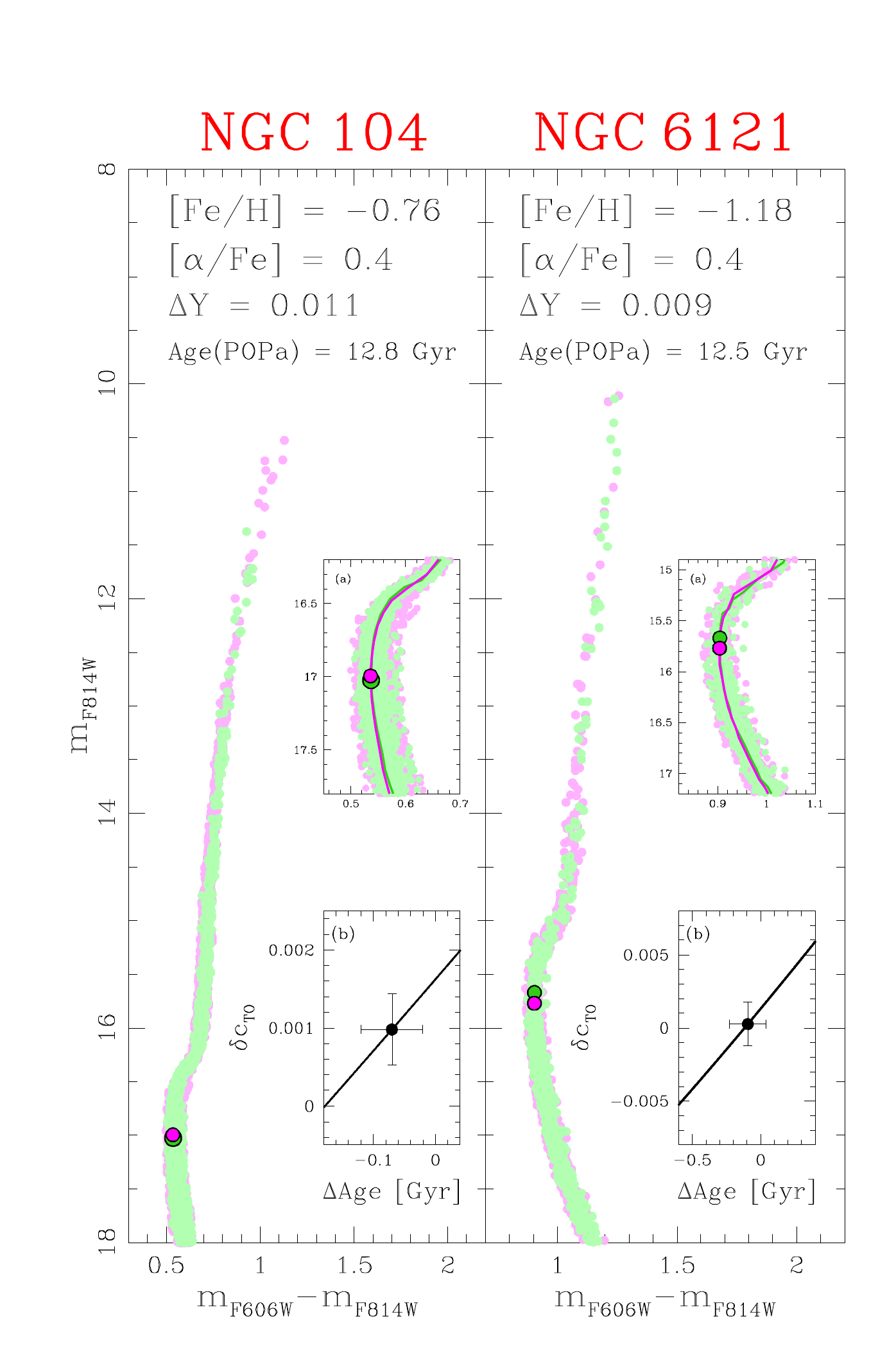}
	\caption{Application of our technique to estimate the relative ages of MPs in NGC\,104 and NGC\,6121. In the observed $m_{\rm F814W}$ versus $m_{\rm F606W}-m_{\rm F814W}$ CMDs POPa (green), POPb (magenta) and their respective MSTOs are plotted. The inserts (a) show a zoomed of the MSTO region. The blue and red fiducial lines are representative of POPa and POPb, respectively. The theoretical models in the insert (b) were obtained assuming the parameters list on the top of the figure.}
	\label{relage1}	
\end{figure*}

\begin{figure*}
\centering
	\includegraphics[trim= 0.8cm 5.5cm 1cm 2.8cm, clip,width=0.7\textwidth]{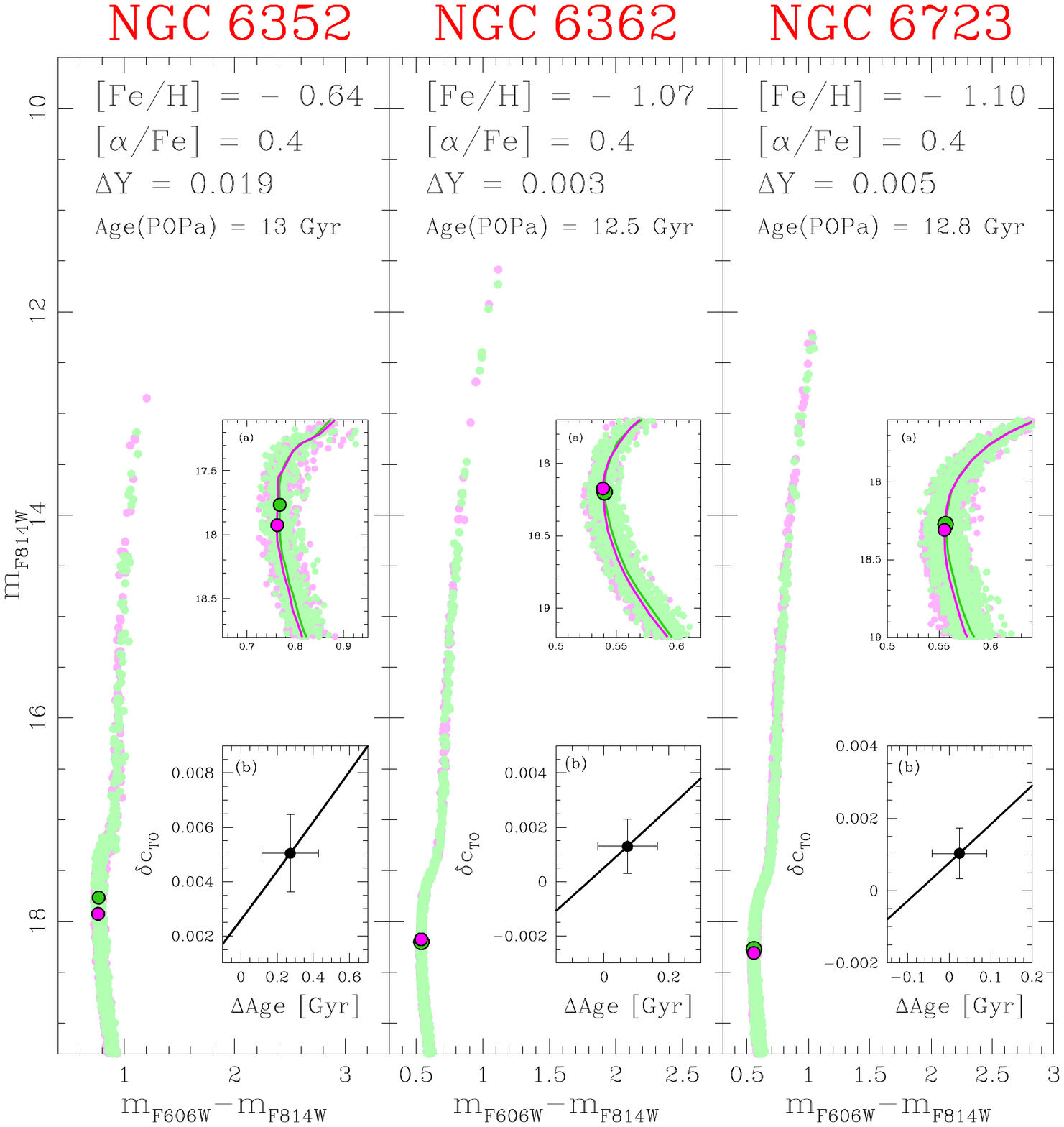}
	\caption{As in Figure \ref{relage1} for the GCs NGC\,6352, NGC\,6362 and NGC\,6723.}
	\label{relage2}	
\end{figure*}

\subsection{Theoretical models}
The procedure adopted to obtain the theoretical models is shown in
Figure \ref{teo}, where NGC\,6362 is taken as an example. \\
For POPa, we considered a set of isochrones from the Dartmouth Stellar 
Evolution Database\footnote{http://stellar.dartmouth.edu/models/} (DSED,
\citealt{2008ApJS..178...89D}) characterized by [Fe/H]$=-1.07$
(\citealt{2017MNRAS.468.1249M}), [$\alpha$/Fe]$=0.4$, primordial helium 
$Y_p =0.2496$ and ages that run from 10 000 to 15 000 Myr, in step of 100 Myr. 
The groups of 2G stars in GCs are helium enriched.
\cite{2018MNRAS.481.5098M} determined the average and the maximum helium 
difference between 2G and 1G stars in a sample of 57 GCs, including our targets.
They estimated a mean $\Delta Y = 0.003$ for NGC\,6362 and we assumed this 
value as helium difference between POPa and POPb of this cluster. 
In order to obtain a set of isochrones with helium content expected for POPb, 
we interpolated the set of isochrones of POPa and a set with same 
value of [Fe/H], [$\alpha$/Fe] and range in age, but with $Y=0.33$.
Panel (a) of Figure \ref{teo} shows three isochrones calculated for an age of 12.5 Gyr, 
[Fe/H]$=-1.07$, [$\alpha$/Fe]$=0.4$, but different helium content: 
the green model (POPa) has primordial helium $Y_p =0.2496$, while 
the red (for POPb) and blue ones have $Y=0.2526$ and $Y=0.33$, respectively. 
We interpolated with a spline each isochrone of both the sets for POPa 
and POPb with a vector having absolute F814W magnitude $0 \leq M_{\rm F814W} \leq 6$, 
whose points are evenly spaced by 0.001 mag. 
For each isochrone of each population, we identify the MS turn-off (MSTO) as the bluer point of the MS. 
In order to obtain a theoretical model appropriate for NGC\,6362, we considered 
for POPa the isochrone having [Fe/H]$=-1.07$, [$ \alpha $/Fe]$=0.4$, $Y_{p}=0.2496$ 
and absolute age 12.5 Gyr \citep{2010ApJ...708..698D}, and for POPb 
the isochrones with the same value of [Fe/H] and [$\alpha$ /Fe], but $Y=0.2526$ and all ages.
We derived the difference in color between the MSTOs, $\delta$c$_{TO}$, 
as the difference between the MSTO color of POPa and POPb. 
In panel (b) of Fig. \ref{teo} we show two cases to clarify this passage. 
The green isochrone represents POPa and corresponds to [Fe/H]$=-1.07$, 
[$\alpha $/Fe]$=0.4$, $Y_{p}=0.2496$ and an absolute age of 12.5 Gyr. 
POPb is represented by the purple and cyan isochrones, which are calculated 
by assuming the same value of [Fe/H], [$\alpha$ /Fe] of POPa, $Y=0.2526$ 
and ages 11 Gyr and 10 Gyr, respectively. The corresponding MSTOs 
of each isochrone are reported as a dot, colored as their isochrone. 
The corresponding difference in color, $\delta$c$_{TO}$, are shown as purple and cyan lines.
We calculated $\Delta$Age subtracting from the MSTO age of POPa, 
i.e the absolute age of 12.5 Gyr, the MSTO age of all the isochrones of POPb. 
The theoretical model $\delta$c$_{TO}$ vs $\Delta$Age is reported in panel (c) of Fig. \ref{teo}.
The purple and cyan points represent the cases explained in panel (b).
In order to obtain a more robust model we interpolated 
these points with a second-order polynomial, where the order was chosen to 
minimize the $\chi^2$ between the polynomial and the observed profile.
In this way, we obtained the black theoretical model.\\
Performing the same procedure and considering the values of [Fe/H] 
(\citealt{2009A&A...505..117C} for NGC\,104, NGC\,6121, NGC\,6723;
 \citealt{2015MNRAS.451..312N} for NGC\,6352; \citealt{2017MNRAS.468.1249M} for NGC\,6362), 
 [$\alpha/$Fe], $\Delta Y$ \citep{2018MNRAS.481.5098M} and absolute age 
 \citep{2010ApJ...708..698D} reported in Table \ref{teomod}, 
we achieved the theoretical models of the other GCs.
A different choice of $\alpha$-enrichment strongly influence the models trend on the RGB, while the position of the MSTO changes insignificantly. Since the sum of light elements is constant, we can assume the same [$\alpha$/Fe] value for POPa and POPb, making the relative MSTO position even more invariant.\\
It is worth noting that the theoretical models are built  assuming a fixed absolute age for POPa 
based on the \cite{2010ApJ...708..698D} estimation, which uncertainties are of 0.5 Gyr.
In Appendix A, we evaluate how the theoretical model is affected by this error, 
and we prove that it does not influence the final relative age result.

%%%%%%%%%%%

\subsection{The relative age}
To obtain the relative age between POPa and POPb within the five GCs 
we compared the theoretical models and the observed $\delta$c$_{TO}$ values, 
as shown in Figures \ref{relage1} and \ref{relage2}.\\
The GC NGC\,6362 in the middle panel of Figure \ref{relage2} is 
taken as an example to explain this step.
We measured the color and the magnitude of the MSTOs of POPa and POPb 
in the $m_{\rm F814W}$ versus $m_{\rm F606W}-m_{\rm F814W}$ CMD.  
The MSTO of each population were obtained using fiducial lines: 
we divided the MS F814W magnitudes regime $17.8 \leq m_{\rm F814W} \leq 19.3$ 
of NGC\,6362  in bins of 0.3 mag. In each bin, we calculated the median magnitude 
and color with a 3$\sigma$-clipping procedure.  The median points were smoothed 
with boxcar averaging filter having a window of 3 points and then were interpolated 
with a spline to obtain the fiducial line.  The MSTO of each population is the bluer 
point of the corresponding fiducial line. We found that the color and F814W magnitude 
of the MSTOs of POPa and POPb of NGC\,6362 are ($0.540\pm 0.001,18.200\pm 0.008$) 
and ($0.539\pm 0.001,18.175\pm 0.009$), respectively.
Furthermore, we calculated the value of $\delta$c$_{TO}$=0.001$\pm$0.001 for NGC\,6362.
The uncertainty on this last value was estimated as the standard deviation from the mean value.
In the $m_{\rm F814W}$ versus $m_{\rm F606W}-m_{\rm F814W}$ CMD of NGC\,6362 
(Figure \ref{relage2}), POPa and POPb are shown in green and magenta and their 
respective MSTOs are reported as filled circles colored as the population they belong to.\\
The MSTO colors and F814W magnitudes and the observed $\delta$c$_{TO}$ 
obtained for the five GCs considered in this work are reported in Table \ref{MSTOGCs}. \\
In order to estimate the relative age between POPa and POPb in NGC\,6362, in the insert 
of Figure \ref{relage2}, we show the observed $\delta$c$_{TO}$ on the theoretical model. 
We concluded that the two main populations of NGC\,6362 have a difference in age 
$\Delta$Age=73$\pm$92 Myr.
The uncertainty associated to this estimation of relative age was obtained interpolating the $\delta$c$_{TO}$ errors with the theoretical model, and it is only due to internal errors.\\
In addition to these internal errors, we analyzed how the helium
uncertainty $\sigma_{\rm \Delta Y}=\pm0.011$ (\citealt{2018MNRAS.481.5098M})
and the metallicity one $\sigma_{\rm [Fe/H]}=\pm0.05$ 
(\citealt{2017MNRAS.468.1249M}) affect the measure of $\Delta$Age.
We produced two additional sets of isochrones for POPb characterized by the same [Fe/H]=-1.07 
and [$\alpha$/Fe]=0.4 of POPa, but with two different helium enhancements: 
$\Delta$Y=0.003+0.011=0.014 and $\Delta$Y=0.003-0.011=-0.008.
We extracted the theoretical models, adopting these two helium enhanced isochrones for 
POPb and the same isochrone as used above for POPa. 
Following the procedure previously outlined and comparing the observed $\delta$c$_{TO}$ 
with these models we found that an uncertainty of $\pm 0.011$ dex in $\Delta$Y  leads to
an uncertainty on $\Delta$Age of $\sigma_{ \Delta Age}(\Delta Y)=\pm 126$ Myr.
Similarly, we calculated the impact of [Fe/H] variations on the estimate of $\Delta$Age.  
In this case, we used for POPb two sets of isochrones with [$\alpha$/Fe]=0.4, enhanced
in helium by $\Delta$Y=0.003 but different metallicity:
[Fe/H]$=-1.07+0.05=-1.02$ and [Fe/H]$=-1.07-0.05=-1.12$.  We found that an
error of $\sigma_{\rm [Fe/H]}$=$\pm$0.05 produce an uncertainty 
$\sigma_{\Delta Age}$([Fe/H]) = 448 Myr. \\
Combining all the sources of uncertainties we can establish an upper
limit for the uncertainty of the relative age between the two main populations within the cluster. 
We found that POPa and POPb of NGC\,6362 are coeval within $\sim 474$\,Myr. \\
We performed the same procedure for the others GCs using the values reported
in Table \ref{teomod}. The final results are summarized in Table \ref{results}.

\begin{table*}
	\centering
	\caption{MPs relative ages and their uncertainties.}
	\label{results}
	\begin{tabular}{lccccc}
          \hline
         \textbf{CLUSTER} & \textbf{$\Delta Age $}&$\sigma _{\Delta Age}$(Internal)&$\sigma _{\Delta Age}$($\Delta$Y) & $\sigma _{\Delta Age}$([Fe/H]) & \textbf{$\sigma _{\Delta Age}$} \\
         &[Myr]&[Myr]&[Myr]&[Myr]&[Myr]\\
          \hline
	NGC\,104 &  -70 & 49 & 74 & 202  &  +220\\
	NGC\,6121  & -77 & 130 & 68 & 157  & +214  \\
	NGC\,6352 & 273 & 150  & 206 & 220 & +336\\
	NGC\,6362  & 73 & 92 & 126 &  448 & +474 \\
	NGC\,6723 & 25 & 65 & 72 & 627 & +634  \\
	\hline	
          \end{tabular}
\end{table*}

%__________________________________________________________________________________
\section{DISCUSSION}
\begin{figure}
	\includegraphics[trim= 0.8cm 5.5cm 1cm 2.8cm, clip,width=0.5\textwidth]{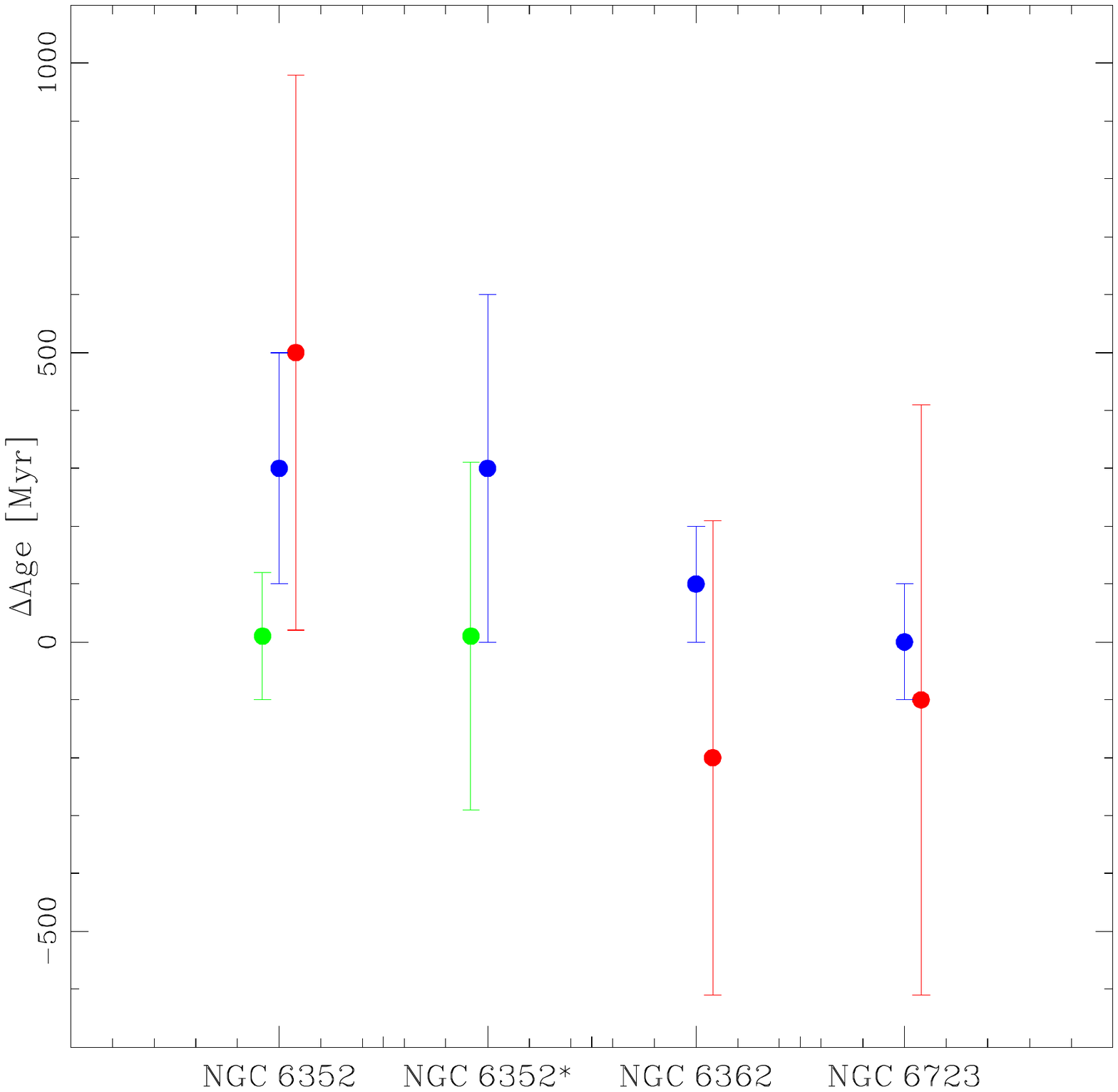}
	\caption{Comparison between this work results (blue) with those obtained by \cite{2015MNRAS.451..312N} (green) and \cite{2020ApJ...891...37O} (red) for NGC\,6352, NGC\,6362 and NGC\,6723.
	The errors bars of NGC\,6352$^*$ include all the uncertainties, while those of the other results consider only the uncertainty on $\Delta$Y.}
	\label{comp}	
\end{figure}

The mechanisms that have brought to the formation of MPs in GCs are
still subject of debate. 
Despite the increasing interest in this topic in the last couple of decades, 
further and deeper analyses are needed. This work aims to introduce a new technique 
that can contribute to put light on this astronomical issue. Indeed, a good relative age 
estimate of MPs leads to clues regarding the formation and the evolution of these last.\\
In literature, most of the works on the relative age of MPs concern GCs with 
a large variation of metallicity between different populations. 
According to \cite{2012A&A...541A..15M}, M\,22 hosts MPs coeval within $
\sim$300\,Myr.  Studying the globular cluster NGC\,2419, 
\cite{2013ApJ...778L..13L} showed that this object hosts two stellar populations characterized 
by a large difference in metallicity and helium abundance. 
They found that the most metal-rich population is younger than 2\,Gyr.
The relative ages of the MPs in NGC\,2808 were analyzed by \cite{2011ApJ...733L..45R}; 
they found that, if 2G stars are helium and metal enhanced by $\Delta$Y=0.03 and $\Delta$Y=0.16, 
respectively, compared to 1G stars, then the 2G
population is $\sim 1.5$\,Gyr younger than the 1G population.\\
Using \textit{HST} data, \cite{2020ApJ...890...38S} estimated an age difference of 550$\pm$410 Myr between the first and the third generations in NGC\,6752. The uncertainty of their result decreases to 400 Myr when the helium enhancement is considered.\\
We compare the results in literature with what we obtained in this work, in cases where the same source of uncertainties were taken into account.\\
\cite{2015MNRAS.451..312N} applied isochrone fitting over synthetic CMDs with $\chi^{2}$ 
calculations to evaluate the relative age of MPs within NGC\,6352.
Assuming [Fe/H]=-0.67, [$\alpha$/Fe]=+0.4 and $\Delta$Y=0.029, they derived an age difference 
of 10$\pm$110 Myr. Considering the same value of metallicity and $\alpha$-enhancement and 
$\Delta$Y=0.019, we estimate $\Delta$Age=273$\pm$150 Myr for NGC\,6352.  
Adopting a difference in [Fe/H] and [$\alpha$/Fe] of 0.02 dex, \cite{2015MNRAS.451..312N} 
found that the two populations are coeval within $\sim$300\,Myr. 
This result is perfectly in agreement with our value of $\sigma _{\Delta Age}$=336 Myr, 
when all the uncertainties are considered.\\
Recently, \cite{2020ApJ...891...37O} used statistical isochrone fitting to estimate the relative age of MPs
 within eight GCs: NGC\,6304, NGC\,6352, NGC\,6362, NGC\,6624, NGC\,6637, NGC\,6652,  
NGC\,6717 and NGC\,6723. They found that the individual MPs are coeval within 500 Myr.
Moreover, they derived a weighted mean age difference of 41$\pm$170 Myr adopting canonical He abundances, 
which reduces to 17$\pm$170 when He-enhancement is taken into account. \\
Considering [Fe/H]=-0.59, [$\alpha$/Fe]=+0.2 and $\Delta$Y=0.027, they derived an age difference 
of 500$\pm$480 Myr for NGC\,6352. This estimation is comparable to our of 273$\pm$150 Myr.\\
Assuming 2G He-enriched by $\Delta$Y=0.004, \cite{2020ApJ...891...37O} found that 
the relative age of MPs in NGC\,6362 is -200$\pm$410 Myr. 
Using $\Delta$Y=0.003, we estimated $\Delta$Age=73$\pm$92 Myr. 
Despite our mean differential age estimation is slightly higher, it agrees within $1\sigma$ 
with the value obtained by \cite{2020ApJ...891...37O}.\\
Finally, we found $\Delta$Age=25$\pm$65 Myr for the MPs in NGC\,6723, which is in agreement 
with the difference in age  -100$\pm$510 Myr obtained by \cite{2020ApJ...891...37O}.\\
In Figure \ref{comp}, this work and literature results for GCs in common are compared. 
We conclude that our new technique leads to consistent result with those in literature and with, on average, smaller error bars.

%_______________________________________________________________________________
\section{CONCLUSIONS}
We developed a new technique to estimate the relative ages of the MPs hosted by GCs.
In this work, we applied the method to five clusters: NGC\,104, NGC\,6121, NGC\,6352, NGC\,6362, NGC\,6723.\\
We used the astro-photometric catalogs released by \cite{2018MNRAS.481.3382N} and 
we selected the well measured stars on the basis of their photometric parameters.\\
A statistical test was used to divide the MPs along the MS of the GCs.
We built the $m_{\rm F336W}-m_{\rm F438W}$ versus $m_{\rm F275W}-m_{\rm F336W}$ 
two-color diagram of SGB and RGB stars in order to divide the stars in the different 
evolutionary phases into the populations which they belong to.
We defined POPa as the 1G stars and POPb as all the successive generations of stars, 
and we considered them as simple stellar populations to estimate their relative age.\\
Considering the values of [Fe/H], [$\alpha$/Fe], Y reported in Tab.\ref{teomod}, 
we built the $\delta$c$_{TO}$ versus $\Delta$Age theoretical model for each GCs.
We defined $\delta$c$_{TO}$ as the difference in color between the MSTO color 
of a fixed isochrone age for POPa and the MSTO of all ages isochrones for POPb 
and the $\Delta$Age as the difference between a fixed age of POPa and all the ages of POPb.
The observed $\delta$c$_{TO}$ was calculated in the $m_{\rm F814W}$ versus 
$m_{\rm F606W}-m_{\rm F814W}$ CMD, and we obtained the relative age between the two main populations.\\
The conclusions drawn from our new technique are:
\begin{itemize}
\item An uncertainty of $\pm$0.5 Gyr (\citealt{2010ApJ...708..698D}) on the absolute age  of POPa 
does not affect the final value of the relative ages (see Appendix A);
\item Combining all the sources of uncertainties (photometry, metallicity and He-enrichment) 
we estimated an upper limit on the relative ages. 
We found that the MPs of NGC\,104 and NGC\,6121 are coeval within 220 Myr and 214 Myr, while those of 
NGC\,6352 , NGC\,6362 and NGC\,6723 are coeval within 336 Myr, 474 Myr and 634 Myr.
\item Within the limits of the errors on relative ages above indicated, the different populations
in the single clusters are coeval.
\end{itemize}
The results obtained with our technique are consistent with those in literature. 
We can affirm that the new method is a good tool to estimate relative ages of MPs within GCs.
Finally, our results turn out new observational evidence to put constrains on the formation of MPs in GCs.
Several theoretical scenarios were proposed to explain this astronomical topic, even if none of them is able to explain all the observational facts obtained in the last years (\citealt{2015MNRAS.454.4197R}).
The most accredited scenarios involve Intermediate Massive AGB stars (\citealt{2002A&A...395...69D}, \citealt{2012MNRAS.423.1521D}), Fast Rotating Massive Stars (\citealt{2007A&A...475..859D}, \citeyear{2007A&A...464.1029D}), Massive Interacting Binaries (\citealt{2009A&A...507L...1D}, \citealt{2013MNRAS.436.2398B}), or Supermassive stars  
(\citealt{2014MNRAS.437L..21D}, \citealt{2015MNRAS.448.3314D}).
According to these scenarios the time scales for the formation of the other populations run from few million years to some hundred Myr.
Unfortunately, with the results obtained in this work we can not discriminate which scenario is the most appropriate to describe the formation of MPs.
Anyway joining our results with the relative ages measured in previous works, we can 
put a strong constraint: the formation of MPs happens on the same time scale for all the normal GCs.

%-------------------------------------
\begin{acknowledgements}
DNa acknowledges the support from the French Centre National d'Etudes Spatiales (CNES).
\end{acknowledgements}

%-------------------------------------------------------------------

\bibliographystyle{aa.bst}
\bibliography{biblio}

\begin{thebibliography}{38}
\expandafter\ifx\csname natexlab\endcsname\relax\def\natexlab#1{#1}\fi

\bibitem[{{Bastian} \& {de Mink}(2009)}]{2009MNRAS.398L..11B}
{Bastian}, N. \& {de Mink}, S.~E. 2009, \mnras, 398, L11

\bibitem[{{Bastian} {et~al.}(2013){Bastian}, {Lamers}, {de Mink}, {Longmore},
  {Goodwin}, \& {Gieles}}]{2013MNRAS.436.2398B}
{Bastian}, N., {Lamers}, H.~J.~G.~L.~M., {de Mink}, S.~E., {et~al.} 2013,
  \mnras, 436, 2398

\bibitem[{{Bellini} {et~al.}(2017){Bellini}, {Anderson}, {Bedin}, {King}, {van
  der Marel}, {Piotto}, \& {Cool}}]{2017ApJ...842....6B}
{Bellini}, A., {Anderson}, J., {Bedin}, L.~R., {et~al.} 2017, \apj, 842, 6

\bibitem[{{Carretta} {et~al.}(2009){Carretta}, {Bragaglia}, {Gratton},
  {Lucatello}, {Catanzaro}, {Leone}, {Bellazzini}, {Claudi}, {D'Orazi},
  {Momany}, {Ortolani}, {Pancino}, {Piotto}, {Recio-Blanco}, \&
  {Sabbi}}]{2009A&A...505..117C}
{Carretta}, E., {Bragaglia}, A., {Gratton}, R.~G., {et~al.} 2009, \aap, 505,
  117

\bibitem[{{D'Antona} {et~al.}(2002){D'Antona}, {Caloi}, {Montalb{\'a}n},
  {Ventura}, \& {Gratton}}]{2002A&A...395...69D}
{D'Antona}, F., {Caloi}, V., {Montalb{\'a}n}, J., {Ventura}, P., \& {Gratton},
  R. 2002, \aap, 395, 69

\bibitem[{{de Mink} {et~al.}(2009){de Mink}, {Pols}, {Langer}, \&
  {Izzard}}]{2009A&A...507L...1D}
{de Mink}, S.~E., {Pols}, O.~R., {Langer}, N., \& {Izzard}, R.~G. 2009, \aap,
  507, L1

\bibitem[{{Decressin} {et~al.}(2007{\natexlab{a}}){Decressin}, {Charbonnel}, \&
  {Meynet}}]{2007A&A...475..859D}
{Decressin}, T., {Charbonnel}, C., \& {Meynet}, G. 2007{\natexlab{a}}, \aap,
  475, 859

\bibitem[{{Decressin} {et~al.}(2007{\natexlab{b}}){Decressin}, {Meynet},
  {Charbonnel}, {Prantzos}, \& {Ekstr{\"o}m}}]{2007A&A...464.1029D}
{Decressin}, T., {Meynet}, G., {Charbonnel}, C., {Prantzos}, N., \&
  {Ekstr{\"o}m}, S. 2007{\natexlab{b}}, \aap, 464, 1029

\bibitem[{{Denissenkov} \& {Hartwick}(2014)}]{2014MNRAS.437L..21D}
{Denissenkov}, P.~A. \& {Hartwick}, F.~D.~A. 2014, \mnras, 437, L21

\bibitem[{{Denissenkov} {et~al.}(2015){Denissenkov}, {VandenBerg}, {Hartwick},
  {Herwig}, {Weiss}, \& {Paxton}}]{2015MNRAS.448.3314D}
{Denissenkov}, P.~A., {VandenBerg}, D.~A., {Hartwick}, F.~D.~A., {et~al.} 2015,
  \mnras, 448, 3314

\bibitem[{{D'Ercole} {et~al.}(2012){D'Ercole}, {D'Antona}, {Carini},
  {Vesperini}, \& {Ventura}}]{2012MNRAS.423.1521D}
{D'Ercole}, A., {D'Antona}, F., {Carini}, R., {Vesperini}, E., \& {Ventura}, P.
  2012, \mnras, 423, 1521

\bibitem[{{Dotter} {et~al.}(2008){Dotter}, {Chaboyer}, {Jevremovi{\'c}},
  {Kostov}, {Baron}, \& {Ferguson}}]{2008ApJS..178...89D}
{Dotter}, A., {Chaboyer}, B., {Jevremovi{\'c}}, D., {et~al.} 2008, \apjs, 178,
  89

\bibitem[{{Dotter} {et~al.}(2010){Dotter}, {Sarajedini}, {Anderson},
  {Aparicio}, {Bedin}, {Chaboyer}, {Majewski}, {Mar{\'{\i}}n-Franch}, {Milone},
  {Paust}, {Piotto}, {Reid}, {Rosenberg}, \& {Siegel}}]{2010ApJ...708..698D}
{Dotter}, A., {Sarajedini}, A., {Anderson}, J., {et~al.} 2010, \apj, 708, 698

\bibitem[{{Gilligan} {et~al.}(2019){Gilligan}, {Chaboyer}, {Cummings},
  {Mackey}, {Cohen}, {Geisler}, {Grocholski}, {Parisi}, {Sarajedini},
  {Ventura}, {Villanova}, {Yang}, \& {Wagner-Kaiser}}]{2019MNRAS.486.5581G}
{Gilligan}, C.~K., {Chaboyer}, B., {Cummings}, J.~D., {et~al.} 2019, \mnras,
  486, 5581

\bibitem[{{Lee} {et~al.}(2013){Lee}, {Han}, {Joo}, {Jang}, {Na}, {Okamoto},
  {Arimoto}, {Lim}, {Kim}, \& {Yoon}}]{2013ApJ...778L..13L}
{Lee}, Y.-W., {Han}, S.-I., {Joo}, S.-J., {et~al.} 2013, \apjl, 778, L13

\bibitem[{{Marino} {et~al.}(2012){Marino}, {Milone}, {Sneden}, {Bergemann},
  {Kraft}, {Wallerstein}, {Cassisi}, {Aparicio}, {Asplund}, {Bedin}, {Hilker},
  {Lind}, {Momany}, {Piotto}, {Roederer}, {Stetson}, \&
  {Zoccali}}]{2012A&A...541A..15M}
{Marino}, A.~F., {Milone}, A.~P., {Sneden}, C., {et~al.} 2012, \aap, 541, A15

\bibitem[{{Martocchia} {et~al.}(2017){Martocchia}, {Bastian}, {Usher},
  {Kozhurina-Platais}, {Niederhofer}, {Cabrera-Ziri}, {Dalessandro},
  {Hollyhead}, {Kacharov}, {Lardo}, {Larsen}, {Mucciarelli}, {Platais},
  {Salaris}, {Cordero}, {Geisler}, {Hilker}, {Li}, \&
  {Mackey}}]{2017MNRAS.468.3150M}
{Martocchia}, S., {Bastian}, N., {Usher}, C., {et~al.} 2017, \mnras, 468, 3150

\bibitem[{{Martocchia} {et~al.}(2019){Martocchia}, {Dalessandro}, {Lardo},
  {Cabrera-Ziri}, {Bastian}, {Kozhurina-Platais}, {Salaris}, {Chantereau},
  {Geisler}, {Hilker}, {Kacharov}, {Larsen}, {Mucciarelli}, {Niederhofer},
  {Platais}, \& {Usher}}]{2019MNRAS.487.5324M}
{Martocchia}, S., {Dalessandro}, E., {Lardo}, C., {et~al.} 2019, \mnras, 487,
  5324

\bibitem[{{Martocchia} {et~al.}(2018){Martocchia}, {Niederhofer},
  {Dalessandro}, {Bastian}, {Kacharov}, {Usher}, {Cabrera-Ziri}, {Lardo},
  {Cassisi}, {Geisler}, {Hilker}, {Hollyhead}, {Kozhurina-Platais}, {Larsen},
  {Mackey}, {Mucciarelli}, {Platais}, \& {Salaris}}]{2018MNRAS.477.4696M}
{Martocchia}, S., {Niederhofer}, F., {Dalessandro}, E., {et~al.} 2018, \mnras,
  477, 4696

\bibitem[{{Massari} {et~al.}(2017){Massari}, {Mucciarelli}, {Dalessandro},
  {Bellazzini}, {Cassisi}, {Fiorentino}, {Ibata}, {Lardo}, \&
  {Salaris}}]{2017MNRAS.468.1249M}
{Massari}, D., {Mucciarelli}, A., {Dalessandro}, E., {et~al.} 2017, \mnras,
  468, 1249

\bibitem[{{Milone} {et~al.}(2013){Milone}, {Marino}, {Piotto}, {Bedin},
  {Anderson}, {Aparicio}, {Bellini}, {Cassisi}, {D'Antona}, {Grundahl},
  {Monelli}, \& {Yong}}]{2013ApJ...767..120M}
{Milone}, A.~P., {Marino}, A.~F., {Piotto}, G., {et~al.} 2013, \apj, 767, 120

\bibitem[{{Milone} {et~al.}(2015){Milone}, {Marino}, {Piotto}, {Bedin},
  {Anderson}, {Renzini}, {King}, {Bellini}, {Brown}, {Cassisi}, {D'Antona},
  {Jerjen}, {Nardiello}, {Salaris}, {Marel}, {Vesperini}, {Yong}, {Aparicio},
  {Sarajedini}, \& {Zoccali}}]{2015MNRAS.447..927M}
{Milone}, A.~P., {Marino}, A.~F., {Piotto}, G., {et~al.} 2015, \mnras, 447, 927

\bibitem[{{Milone} {et~al.}(2018){Milone}, {Marino}, {Renzini}, {D'Antona},
  {Anderson}, {Barbuy}, {Bedin}, {Bellini}, {Brown}, {Cassisi}, {Cordoni},
  {Lagioia}, {Nardiello}, {Ortolani}, {Piotto}, {Sarajedini}, {Tailo}, {van der
  Marel}, \& {Vesperini}}]{2018MNRAS.481.5098M}
{Milone}, A.~P., {Marino}, A.~F., {Renzini}, A., {et~al.} 2018, \mnras, 481,
  5098

\bibitem[{{Milone} {et~al.}(2012{\natexlab{a}}){Milone}, {Piotto}, {Bedin},
  {Aparicio}, {Anderson}, {Sarajedini}, {Marino}, {Moretti}, {Davies},
  {Chaboyer}, {Dotter}, {Hempel}, {Mar{\'{\i}}n-Franch}, {Majewski}, {Paust},
  {Reid}, {Rosenberg}, \& {Siegel}}]{2012A&A...540A..16M}
{Milone}, A.~P., {Piotto}, G., {Bedin}, L.~R., {et~al.} 2012{\natexlab{a}},
  \aap, 540, A16

\bibitem[{{Milone} {et~al.}(2012{\natexlab{b}}){Milone}, {Piotto}, {Bedin},
  {Marino}, {Momany}, \& {Villanova}}]{2012MSAIS..19..173M}
{Milone}, A.~P., {Piotto}, G., {Bedin}, L.~R., {et~al.} 2012{\natexlab{b}},
  Memorie della Societa Astronomica Italiana Supplementi, 19, 173

\bibitem[{{Nardiello} {et~al.}(2018{\natexlab{a}}){Nardiello}, {Libralato},
  {Piotto}, {Anderson}, {Bellini}, {Aparicio}, {Bedin}, {Cassisi}, {Granata},
  {King}, {Lucertini}, {Marino}, {Milone}, {Ortolani}, {Platais}, \& {van der
  Marel}}]{2018MNRAS.481.3382N}
{Nardiello}, D., {Libralato}, M., {Piotto}, G., {et~al.} 2018{\natexlab{a}},
  \mnras, 481, 3382

\bibitem[{{Nardiello} {et~al.}(2018{\natexlab{b}}){Nardiello}, {Milone},
  {Piotto}, {Anderson}, {Bedin}, {Bellini}, {Cassisi}, {Libralato}, \&
  {Marino}}]{2018MNRAS.477.2004N}
{Nardiello}, D., {Milone}, A.~P., {Piotto}, G., {et~al.} 2018{\natexlab{b}},
  \mnras, 477, 2004

\bibitem[{{Nardiello} {et~al.}(2015){Nardiello}, {Piotto}, {Milone}, {Marino},
  {Bedin}, {Anderson}, {Aparicio}, {Bellini}, {Cassisi}, {D'Antona}, {Hidalgo},
  {Ortolani}, {Pietrinferni}, {Renzini}, {Salaris}, {Marel}, \&
  {Vesperini}}]{2015MNRAS.451..312N}
{Nardiello}, D., {Piotto}, G., {Milone}, A.~P., {et~al.} 2015, \mnras, 451, 312

\bibitem[{{Nardiello} {et~al.}(2019){Nardiello}, {Piotto}, {Milone}, {Rich},
  {Cassisi}, {Bedin}, {Bellini}, \& {Renzini}}]{2019MNRAS.485.3076N}
{Nardiello}, D., {Piotto}, G., {Milone}, A.~P., {et~al.} 2019, \mnras, 485,
  3076

\bibitem[{{Oliveira} {et~al.}(2020){Oliveira}, {Souza}, {Kerber}, {Barbuy},
  {Ortolani}, {Piotto}, {Nardiello}, {P{\'e}rez-Villegas}, {Maia}, {Bica},
  {Cassisi}, {D'Antona}, {Lagioia}, {Libralato}, {Milone}, {Anderson},
  {Aparicio}, {Bedin}, {Brown}, {King}, {Marino}, {Pietrinferni}, {Renzini},
  {Sarajedini}, {van der Marel}, \& {Vesperini}}]{2020ApJ...891...37O}
{Oliveira}, R.~A.~P., {Souza}, S.~O., {Kerber}, L.~O., {et~al.} 2020, \apj,
  891, 37

\bibitem[{{Piotto} {et~al.}(2015){Piotto}, {Milone}, {Bedin}, {Anderson},
  {King}, {Marino}, {Nardiello}, {Aparicio}, {Barbuy}, {Bellini}, {Brown},
  {Cassisi}, {Cool}, {Cunial}, {Dalessandro}, {D'Antona}, {Ferraro}, {Hidalgo},
  {Lanzoni}, {Monelli}, {Ortolani}, {Renzini}, {Salaris}, {Sarajedini}, {van
  der Marel}, {Vesperini}, \& {Zoccali}}]{2015AJ....149...91P}
{Piotto}, G., {Milone}, A.~P., {Bedin}, L.~R., {et~al.} 2015, \aj, 149, 91

\bibitem[{{Renzini} {et~al.}(2015){Renzini}, {D'Antona}, {Cassisi}, {King},
  {Milone}, {Ventura}, {Anderson}, {Bedin}, {Bellini}, {Brown}, {Piotto}, {van
  der Marel}, {Barbuy}, {Dalessandro}, {Hidalgo}, {Marino}, {Ortolani},
  {Salaris}, \& {Sarajedini}}]{2015MNRAS.454.4197R}
{Renzini}, A., {D'Antona}, F., {Cassisi}, S., {et~al.} 2015, \mnras, 454, 4197

\bibitem[{{Roh} {et~al.}(2011){Roh}, {Lee}, {Joo}, {Han}, {Sohn}, \&
  {Lee}}]{2011ApJ...733L..45R}
{Roh}, D.-G., {Lee}, Y.-W., {Joo}, S.-J., {et~al.} 2011, \apjl, 733, L45

\bibitem[{{Rosenberg} {et~al.}(1999){Rosenberg}, {Saviane}, {Piotto}, \&
  {Aparicio}}]{1999AJ....118.2306R}
{Rosenberg}, A., {Saviane}, I., {Piotto}, G., \& {Aparicio}, A. 1999, \aj, 118,
  2306

\bibitem[{{Saracino} {et~al.}(2019){Saracino}, {Bastian}, {Kozhurina-Platais},
  {Cabrera-Ziri}, {Dalessandro}, {Kacharov}, {Lardo}, {Larsen}, {Mucciarelli},
  {Platais}, \& {Salaris}}]{2019MNRAS.489L..97S}
{Saracino}, S., {Bastian}, N., {Kozhurina-Platais}, V., {et~al.} 2019, \mnras,
  489, L97

\bibitem[{{Saracino} {et~al.}(2020){Saracino}, {Martocchia}, {Bastian},
  {Kozhurina-Platais}, {Chantereau}, {Salaris}, {Cabrera-Ziri}, {Dalessandro},
  {Kacharov}, {Lardo}, {Larsen}, \& {Platais}}]{2020MNRAS.493.6060S}
{Saracino}, S., {Martocchia}, S., {Bastian}, N., {et~al.} 2020, \mnras, 493,
  6060

\bibitem[{{Sarajedini} {et~al.}(2006){Sarajedini}, {Anderson}, {Aparicio},
  {Bedin}, {Chaboyer}, {Dotter}, {Hempel}, {King}, {Majewski}, {Marin-Franch},
  {Milone}, {Paust}, {Piotto}, {Reid}, {Rosenberg}, \&
  {Siegel}}]{2006AAS...20910009S}
{Sarajedini}, A., {Anderson}, J., {Aparicio}, A., {et~al.} 2006, in Bulletin of
  the American Astronomical Society, Vol.~38, American Astronomical Society
  Meeting Abstracts, 1044

\bibitem[{{Souza} {et~al.}(2020){Souza}, {Kerber}, {Barbuy},
  {P{\'e}rez-Villegas}, {Oliveira}, \& {Nardiello}}]{2020ApJ...890...38S}
{Souza}, S.~O., {Kerber}, L.~O., {Barbuy}, B., {et~al.} 2020, \apj, 890, 38

\end{thebibliography}

 %___________________________________________________________
 
\begin{appendix}
\section{Theoretical model}
\begin{figure}
\centering
	\includegraphics[trim= 0.7cm 5.5cm 1cm 2.5cm, clip, width=0.5\textwidth]{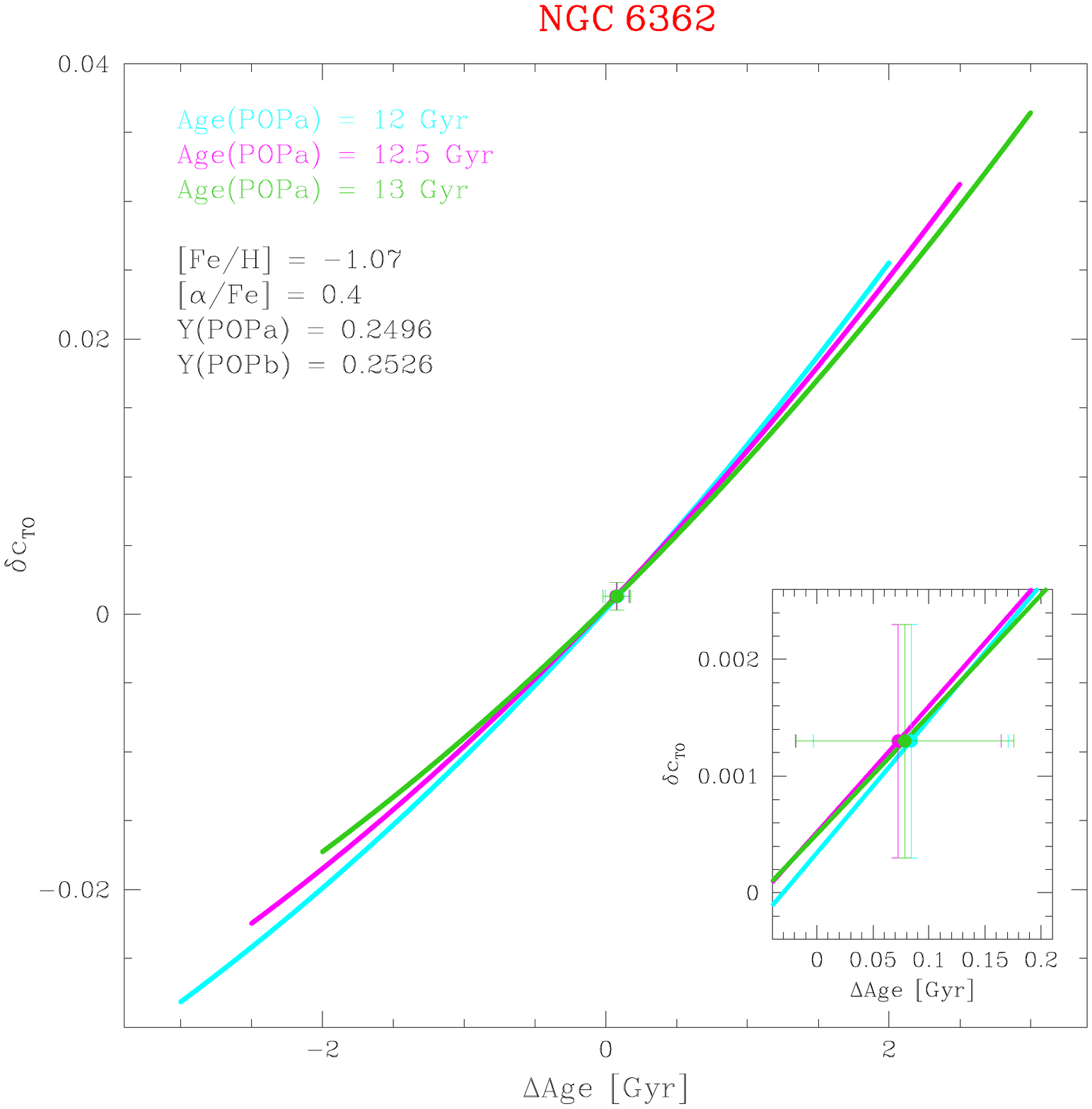}
	\caption{Application of the new technique to estimate the relative age of MPs within NGC\,6362. The theoretical models were obtained considering POPa ages of 12 Gyr (cyan), 12.5 Gyr (magenta) and 13 Gyr (green). A better comparison between the three models is shown in the insert.}
	\label{app}	
\end{figure}

\begin{figure}
\centering
	\includegraphics[trim= 0.7cm 5.5cm 1cm 2.5cm, clip,width=0.5\textwidth]{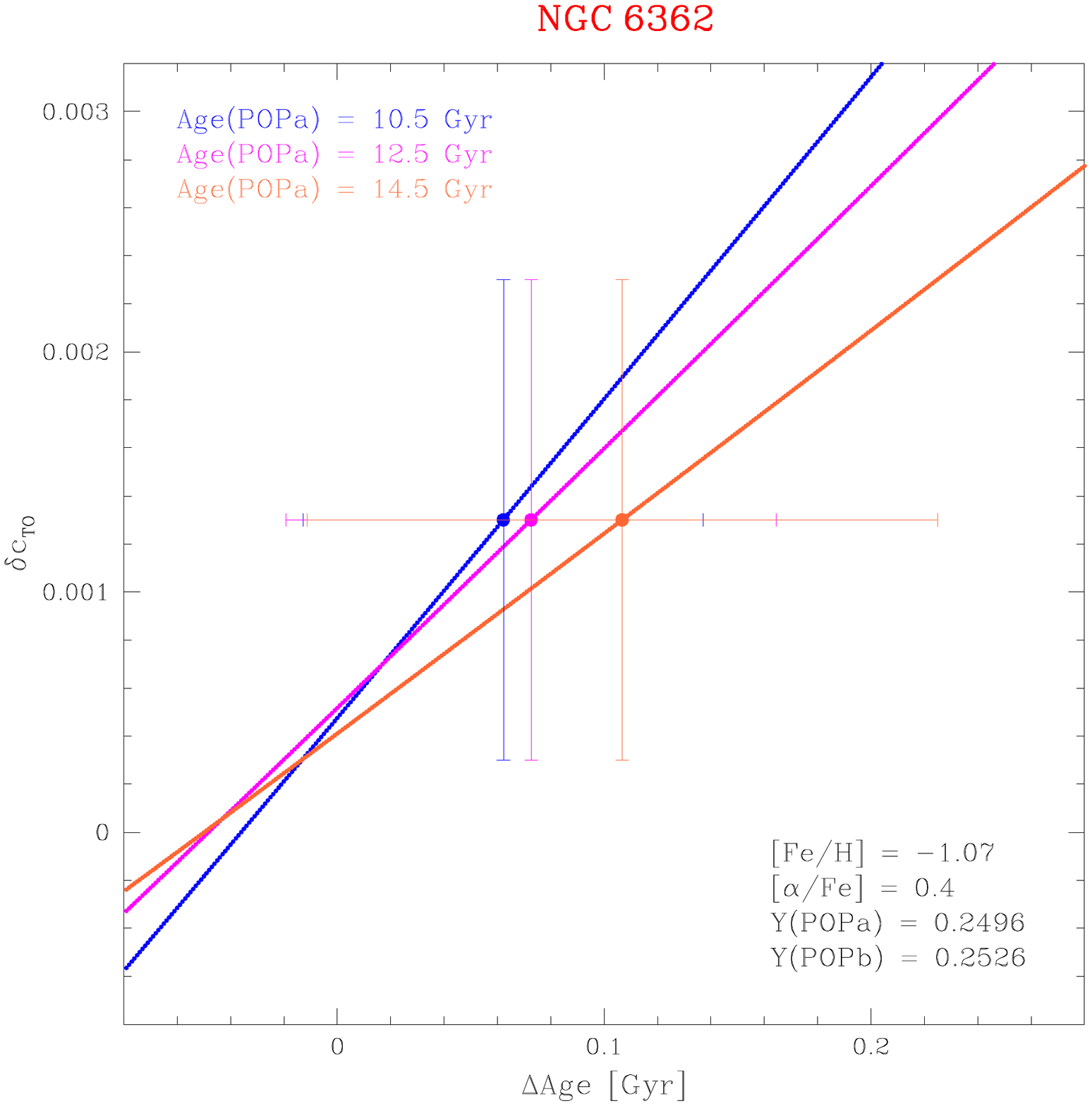}
	\caption{Application of the new technique to estimate the relative age of MPs within NGC\,6362. The theoretical models were obtained considering POPa ages of 10.5 Gyr (blue), 12.5 Gyr (magenta) and 14.5 Gyr (orange).}
	\label{app1}	
\end{figure}

The relative age values are given comparing the observed $\delta c_{TO}$ and the theoretical models. 
The theoretical models are based on an assumption on the absolute age of POPa.
In this work, we adopted the ages obtained by \cite{2010ApJ...708..698D}, 
with on average an uncertainty of $\pm$0.5 Gyr.
This appendix aims to investigate how the error on the POPa age affects the theoretical model 
and, consequently, the relative age estimate.\\
With this purpose, we take the GC NGC\,6362 as example.
In this cluster, we assumed for POPa an isochrone characterized by [Fe/H]=-1.07 (\citealt{2017MNRAS.468.1249M}), 
[$\alpha$/Fe]=+0.4, primordial helium Y=0.2496 and age 12.5 Gyr (\citealt{2010ApJ...708..698D}). 
On the other hand, the set of isochrones for POPb has same [Fe/H] and [$\alpha$/Fe], 
He-enrichment $\Delta$Y=0.003 (\citealt{2018MNRAS.481.5098M}) and ages that 
run from 10 000 Myr to 15 000 Myr, in step of 100 Myr. 
As described in section 4.1 and Figure \ref{teo}, we derived the theoretical model for this GC, 
as shown in Figure \ref{app}  with the color magenta. 
Reporting the observed $\delta c_{TO}$ on this profile, we obtained a relative age of 73 $\pm$ 92 Myr.
The same Figure displays the theoretical models obtained considering an age of 12 Gyr (cyan) and 13 Gyr (green) for POPa. 
As  shown in the insert, there is a negligible difference among the three models in the region where 
the observed $\delta c_{TO}$ lies.  Indeed, the relative ages found considering the cyan and green models 
are 84 $\pm$ 87 Myr and 78 $\pm$ 97 Myr, respectively.
We conclude that an uncertainty of $\pm$0.5 Gyr on the absolute age 
of POPa does not affect the final relative age result.\\
Furthermore, we performed a deeper analysis to evaluate how the choice of POPa age affects the theoretical model. 
We applied the method for NGC\,6362 considering POPa ages of 10.5 Gyr (blue) and 14.5 Gyr (orange), 
as shown in Figure \ref{app1}. From these cases, we estimated relative ages of 62 $\pm$ 75 Myr and 107 $\pm$ 118 Myr, respectively. 
Despite the inclination variation of the theoretical model, the final results are consistent within error bars.
We can affirm that a different choice of POPa age within $\pm$2Gyr, which is the typical range of GC ages, 
does not affect the relative age value estimated with our technique.

\end{appendix}

\end{document}